\documentclass[12pt]{article}
\usepackage{amsmath}
\usepackage{amssymb}
\usepackage{amsfonts}
\usepackage{amscd}
\usepackage[all]{xy}

\title{Families of 2D superintegrable  anisotropic Dunkl oscillators and algebraic derivation of  their spectrum}

\author{Phillip S. Isaac$^{1,}$\thanks{Electronic address: psi@maths.uq.edu.au} \ and Ian Marquette$^{1,}$\thanks{Electronic address:  i.marquette@uq.edu.au}\\
{\small\sl $^1$ School of Mathematics and Physics, The University of Queensland,}\\
{\small \sl Brisbane, QLD 4072, Australia}
}
\date{ }
\begin{document}
\baselineskip=22pt plus 1pt minus 1pt
%%%%%%%%%%%%%%%%%%%%%%%%%%%%%%%%%%%%%%%%%%%%%%%%%%%%%%%%%%
\maketitle

\begin{abstract}
We generalise the construction of integrals of motion for quantum superintegrable
models and the deformed oscillator algebra approach. This is presented in the context of 1D systems admitting ladder
operators satisfying a parabosonic algebra involving reflection operators and more generally
$c_{\lambda}$ extended oscillator algebras with grading. We apply the construction on
two-dimensional $c_{\lambda}$ oscillators. We also introduce two new superintegrable Hamiltonians
that are the anisotropic Dunkl and the singular Dunkl oscillators. We construct the integrals and
using this extended approach of the Daskaloyannis method with grading and we present an algebraic
derivation of the energy spectrum of the two models from the finite dimensional unitary
representations and show how their spectrum divides into different sectors and relates to the
physical spectrum.
\end{abstract}

\vspace{0.5cm}


 
\newpage
%
%========================================================================
%
\section{INTRODUCTION}

It is well known that the deformed oscillator algebras approach (also referred to as the
Daskaloyannis method) may be used to obtain finite dimensional unitary representations of finitely
generated polynomial algebras involving three generators, and to obtain algebraic derivations of the
energy spectrum of various superintegrable systems \cite{bon99,das01,isa14}. Some examples of higher
rank quadratic algebras have also been considered \cite{hoq15a,hoq15b,tan11}. Some of the algebras were
direct sums of Lie algebras and quadratic algebras with structure constants involving Casimir
operators of some higher rank Lie algebras, but central element to the generators of the quadratic
algebra. Other cases of quadratic algebras have been shown to have a more complicated embedded
structure. In addition, methods to generate polynomial algebras and integrals of motion of systems
from ladder operators satisfying polynomial Heisenberg algebras have been used \cite{mar13c}. These
methods have been extended to the case of integrals constructed with combinations of various types of
supercharges, ladder and shift operators \cite{mar10,kal1,kal2,cal1,cal2}. \par
%
%---
%
One-dimensional quantum systems involving the Dunkl operator or reflection operators were studied in
the context of Calogero models and supersymmetric quantum mechanics
\cite{dun1,dun2,muk1,now1,ply1,ply2,sut1}. In recent years, a 2D superintegrable system involving
reflection operators has been discovered and studied \cite{pos1} and a series of papers introduced
isotropic 2D and 3D Dunkl and singular Dunkl oscillators and a particular case of anisotropic with
frequency $2:1$ \cite{pos1,gen1,gen2,gen3,gen4,gen6}. These works made interesting
connections between the obtained quadratic algebras and orthogonal polynomials of various types such as
Jacobi-Dunkl polynomials and dual-1 Hahn polynomials. Some works on angular momentum involving such
operators and the related algebra were also undertaken recently \cite{fei1}. The 1D Dunkl and singular
oscillators also admit ladder operators involving reflection operators and from an algebraic
perspective they comprise a particular case among the class of $c_{\lambda}$ extended oscillators that
involve generators of the cyclic group \cite{bec1,que1,vas1}.\par
%
%---
%

The purpose of this paper is to extend the construction of integrals for 2D Hamiltonians from ladder
operators satisfying a $c_{\lambda}$ extended oscillator algebra for the 1D components of the
Hamiltonian, obtain the corresponding polynomial algebras and demonstrate how the deformed
oscillator algebras approach can be applied to obtain the spectrum. The other main objective is to
introduce two new families of 2D superintegrable systems that are anisotropic generalizations of the
2D Dunkl and singular Dunkl oscillators, present their integrals, polynomial algebras, the
realizations as deformed oscillator algebras and obtain the spectrum.\par
%
%---
%

In Section 2, we present a construction of integrals of motion using ladder operators satisfying a
$c_{\lambda}$ extended oscillator algebra and present the finitely generated polynomial algebras. In
Section 3, we study an anisotropic generalization of Dunkl and singular Dunkl oscillator and present
algebraic derivation of the energy spectrum from the finite-dimensional unitary representation. In
Section 4, we make a comparison with the physical spectrum obtained via separation of variables in
Cartesian coordinates. \par
%
%---
%

\section{Daskaloyannis approach with grading}

We consider Hamiltonians built from two commuting parts
\begin{equation*}
H=H_{1}+H_{2},
\end{equation*}
i.e. such that $[H_{1},H_{2}]=0$, with each part $H_k$ associated to ladder operators $a_k,$
$a_k^\dagger$ satisfying {\em deformed} $c_{\lambda}$-extended oscillator algebra commutation relations \cite{que1}.
Specifically, for $k=1,2$, we let $T_k$ denote generators of the cyclic group of order $\lambda_k$, satisfying
$T_{k}^{\lambda_{k}}=I$, with
$C_{\lambda_{k}}=\mbox{span}\{I,T_{k},T_{k}^{2},\ldots,T_{k}^{\lambda_{k}-1}\}$ being the
corresponding group
algebra such that 
\begin{equation}
a_{k}^{\dagger}T_{k}=q_k^{-1}T_{k}a_k^{\dagger},
\label{basicequ}
\end{equation}
where we have set $q_k =
e^{\frac{i2\pi}{\lambda_k}}$. 
An immediate consequence of this relation is the identity
\begin{equation}
a_{k}^{\dagger}T_{k}^\ell=q_k^{-\ell} T_{k}^\ell a^{\dagger},\ \
\ell=1,2,\ldots,\lambda_k-1.
\label{note1star}
\end{equation}
We furthermore define $T_k^\dagger = T_k^{\lambda_k-1}$, from which it easily follows that
$$
\left(T_k^\ell\right)^\dagger = T_k^{\lambda_k-\ell},\ \ \ell=1,2,\ldots,\lambda_k-1. 
$$
Taking the hermitian conjugate of equation (\ref{note1star}) then
leads to
$$
T_k^\ell a_k = q_k^{-\ell} a_kT_k^\ell, \ \ \ell=1,2,\ldots,\lambda_k-1.
$$
In fact it is straightforward to establish the more general relation
\begin{equation}
a_k^NT_k^m = q_k^{mN}T_k^ma_k^N.
\label{note4doubledagger}
\end{equation}

We now impose the relations
\begin{equation}
[H_{k},a_{k}]
=-a_{k} \left(\alpha_{k,0} I + \sum_{m=1}^{\lambda_{k}-1}\alpha_{k,m}T_{k}^{m}\right).
%= -a_{k}(\alpha_{k,0} I + \sum_{\mu=1}^{\lambda_{k}-1}\delta_{k,\mu}P_{k,\mu}).
\label{cextoscalg}
\end{equation}
Taking 
\begin{equation}
\beta_{k,m}=\bar{\alpha}_{k,\lambda_{k}-m}q_k^m,
\ \ m=1,2,\ldots,\lambda_{k}-1, \ \ \beta_{k,0}=\bar{\alpha}_{k,0}
\label{coeffrels}
\end{equation}
(using the overbar to denote
complex conjugation), the hermitian conjugate of
(\ref{cextoscalg}) gives 
\begin{equation}
[H_{k},a_{k}^{\dagger}]=a_{k}^{\dagger}\left(\beta_{k,0} I +
\sum_{m=1}^{\lambda_{k}-1}\beta_{k,m}T_{k}^{m}\right).
\label{cond}
\end{equation}
If we further impose 
\begin{equation}
\alpha_{k,0}=\bar{\alpha}_{k,0} \mbox{ and }
\alpha_{k,m}=\bar{\alpha}_{k,\lambda_{k}-m},
\label{conjcond}
\end{equation}
then we may deduce 
\begin{equation}
[ H_{k},a_k^{\dagger}a_{k}]=0.
\label{aadaggercomm}
\end{equation}
By contrast to our approach, we note that in the work of Quesne and Vansteenkiste \cite{que1}, the
authors set
$$
[a,a^\dagger] = 1+\sum_{\mu=1}^{\lambda -1}\kappa_\mu T^\mu,
$$
and impose the condition $\bar{\kappa}_\mu = \kappa_{\lambda-\mu}.$ The condition (\ref{conjcond})
could be considered an analogue of this condition.

Using the relations (\ref{cextoscalg}) and (\ref{note4doubledagger}) we may easily deduce by
induction that
\begin{equation}
[H_k,a_k^N] = -\left( N\alpha_{k,0}I + \sum_{m=1}^{\lambda_k-1}\sum_{j=1}^Nq_k^{jm}\alpha_{k,m}T_k^m
\right)a_k^N.
\label{page4prop}
\end{equation}
Furthermore, for the primitive $\lambda_k$-th root of unity $q_k$ and for any
$m=1,2,\ldots,\lambda_k-1$, we have
\begin{equation}
\sum_{j=1}^{\lambda_k}q_k^{jm} = 0.
\label{page5lemma}
\end{equation}
Setting $N=\lambda_k$ in equation (\ref{page4prop}) and applying (\ref{page5lemma}) leads to the commutation
relation
$$
[H_k,a_k^{\lambda_k}] = -\lambda_k\alpha_{k,0}a_k^{\lambda_k}.
$$
In a similar way, we may also deduce the relation
$$
[H_k,(a_k^\dagger)^{\lambda_k}] = \lambda_k\alpha_{k,0}(a_k^\dagger)^{\lambda_k}.
$$
We also note, from (\ref{note4doubledagger}), that 
$$
a_k^{\lambda_k}T_k = T_k a_k^{\lambda_k}.
$$
We may express equation (\ref{cond}) in the form
$$
H_ka_k^\dagger=a_k^\dagger\left( H_k+\sum_{m=0}^{\lambda_k-1}\beta_{k,m}T_k^m \right),
$$
and similarly write equation (\ref{cextoscalg}) as
$$
a_kH_k = \left( H_k + \sum_{m=0}^{\lambda_k-1}\beta_{k,m}T_k^m \right) a_k,
$$
where we have made use of (\ref{coeffrels}) and (\ref{conjcond}) above. We then deduce that for any
analytic function $f$, we have the relations
\begin{align*}
f(H_k)a_k^\dagger &= a_k^\dagger f\left( H_k+\sum_{m=0}^{\lambda_k-1}\beta_{k,m}T_k^m \right),\\
a_kf(H_k) &= f\left( H_k + \sum_{m=0}^{\lambda_k-1}\beta_{k,m}T_k^m \right) a_k.
\end{align*}
This indicates that if we set the analytic function $f$ as
$$
a_k^\dagger a_k = f(H_k),
$$
then we may consistently set
$$
a_ka_k^\dagger = f\left( H_k + \sum_{m=0}^{\lambda_k-1}\beta_{k,m}T_k^m \right).
$$
Even though this is justified by equation (\ref{aadaggercomm}), it tuns out that we may 
consistently impose the condition that the $T_k$ also
commute with $H_k$. Consider the following argument.

We start by investigating the commutator of both sides of the relation (\ref{basicequ}). Using the
standard derivation rule, we have
$$
[H_k,a_k^\dagger T_k] = a_k^\dagger [H_k,T_k] + a_k^\dagger \sum_{m=0}^{\lambda_k-1}
\beta_{k,m}T_k^{m+1},
$$
and
$$
[H_k,q_k^{-1}T_ka_k^\dagger] = a_k^\dagger \sum_{m=0}^{\lambda_k-1} \beta_{k,m}T_k^{m+1} +
q_k^{-1}[H_k,T_k]a_k^\dagger.
$$
Equating both expressions gives
$$
a_k^\dagger [H_k,T_k] = q_k^{-1}[H_k,T_k]a_k^\dagger.
$$
Based on the form of (\ref{basicequ}), we make the assumption that
$$
[H_k,T_k] = \phi_k T_k,
$$
where $\phi_k$ is some constant of proportionality. This then implies that
$$
H_kT_k = T_k(H_k+\phi_k),
$$
which may easily be extended to powers of $T_k$, namely
$$
H_kT_k^m = T_k^m(H_k+m\phi_k).
$$
Setting $m=\lambda_k$, however, leads to the conclusion that $\phi_k=0$, and therefore we have
\begin{equation}
[H_k,T_k]=0.
\label{HTcommrels}
\end{equation}

The outcome of this discussion is that we can impose the relations
\begin{equation}
a_k^\dagger a_k = \sum_{m=0}^{\lambda_k-1}Q_{k,m}(H_k)T_k^m
\label{extend1}
\end{equation} 
and
\begin{equation}
a_ka_k^\dagger = \sum_{m=0}^{\lambda_k-1}Q_{k,m}\left(H_k+\sum_{n=0}^{\lambda_k-1}\beta_{k,n}T_k^n
\right)T_k^m,
\label{extend2}
\end{equation}
where the $Q_{k,m}$ are some specified analytic functions.

% For $k=1,2$ with $T_{1}^{\lambda_{1}}=I$ and $T_{2}^{\lambda_{2}}=I$ are generator of the cyclic
% group of order $\lambda_{1}$ and $\lambda_{2}$,
% $C_{\lambda_{1}}=\{I,T_{1},T_{1}^{2},...,T_{1}^{\lambda_{1}-1}\}$ and
% $C_{\lambda_{2}}=\{I,T_{2},T_{2}^{2},...,T_{2}^{\lambda_{2}-1}\}$ such that
% $a_{k}^{\dagger}T_{k}=e^{-\frac{i2\pi}{\lambda_{k}}}T_{k}a^{\dagger}$.
% 
% There exist other generators 
% \begin{equation}
% T_{k}^{\nu}=\sum_{\mu=0}^{\lambda_{k}-1}e^{\frac{i 2 \pi \mu \nu}{\lambda_{k}}}P_{k,\mu} ,\quad a_{k}^{\dagger}P_{k,\mu}=P_{k,\mu+1}a_{k}^{\dagger}
% \end{equation}
% which satisfy
% 
% 
% 
% 
% 
% 
% \begin{equation}
% a_{k}^{\dagger}a_{k}=Q_{k,0}(H_{k}) I+\sum_{i=1}^{\lambda_{k}-1}S_{k,i}(H_{k})T_{k}^{i}= Q_{k,0}(H_{k}) I + \sum_{i=1}^{\lambda_{k}-1}Q_{k,\mu}(H_{k})P_{k,\mu}
% \end{equation}
% \begin{equation}
% a_{k}a_{k}^{\dagger}=Q_{k,0}(H_{1}+\alpha_{k,0} I + \sum_{\mu}^{\lambda_{k}-1}\delta_{k,\mu}P_{k,\mu}) I + \sum_{i=1}^{\lambda_{k}-1}Q_{k,\mu}( H_{k}+\alpha_{k,0} I + \sum_{\mu}^{\lambda_{k}-1}\delta_{k,\mu}P_{k,\mu}   )P_{k,\mu-1} 
% \end{equation}
% 
Now introduce the new operators % such that $[B_{k},P_{k,\mu}]=0$
\begin{equation}
B_{k}=a_{k}^{ \lambda_{k}},\quad 
B_{k}^{\dagger}=\left( a_{k}^{\dagger} \right)^{\lambda_{k}}.
\label{Bdef}
\end{equation}
We immediately have
\begin{equation}
[H_{k},B_{k}]=-\gamma_k B_k, \quad [H_k,B_k^\dagger] = \gamma_k B_k^\dagger,
\label{Rrels}
\end{equation}
where we have set $\gamma_k = \lambda_k\alpha_{k,0},$ which must be a real number (from
(\ref{conjcond})).
This then leads to 
\begin{equation*}
B_{k}^{\dagger}B_{k}=G_{k}(H_{k}),\quad B_{k}B_{k}^{\dagger}=G_{k}\left(H_{k}+\gamma_k\right),
\end{equation*}
where
$$
G_k(H_k) = \prod_{\ell=1}^{\lambda_k} Z_{\ell,\lambda_k}^{(k)}(H_k,T_k).
$$
Here, for convenience, we have set
$$
Z_{\ell,\lambda_k}^{(k)}(H_k,T_k) = \sum_{m=0}^{\lambda_k-1} q_k^{\ell m}Q_{k,m}\left( H_k -
\sum_{n=0}^{\lambda_k-1}\sum_{j=0}^{\lambda_k-1-\ell}\beta_nq_k^{-jn}T_k^n \right)T_k^m,
$$
and in particular,
$$
Z_{\lambda_k,\lambda_k}^{(k)}(H_k,T_k) = \sum_{m=0}^{\lambda_k-1}Q_{k,m}(H_k)T_k^m.
$$
The generators of the cyclic group are now central elements of the finitely generated polynomial
algebra. We introduce the following generators
\begin{equation}
I_{-}=(B_{1})^{n_{1}} (B_{2}^{\dagger})^{n_{2}},\quad I_{+}=(B_{1}^{\dagger})^{n_{1}}
(B_{2})^{n_{2}},\quad K=\frac{1}{2\gamma}(H_{1}-H_{2}),
\label{defI}
\end{equation}
with constraints $n_{1}\gamma_{1}=n_{2}\gamma_{2}=\gamma$. 
%We choose $n_{1}=\gamma_{1}$,$n_{2}=\gamma_{1}$ and $\gamma=\gamma_{1}\gamma_{2}$.
These generate the following polynomial algebra
\begin{equation*}
[K,I_{\pm}]=\pm I_{\pm},
\end{equation*}
\begin{equation*}
[I_{-},I_{+}] = S_{n_{1},n_{2}}(K+1,H)-S_{n_{1},n_{2}}(K,H),
\end{equation*}
where
\begin{equation}
S_{n_{1},n_{2}}(K,H)=\prod_{i=1}^{n_{1}}G_{1}\left(\frac{1}{2}H+\gamma K
-(n_{1}-i)\gamma_{1}\right)\prod_{j=1}^{n_{2}}G_{2}\left(\frac{1}{2}H-\gamma K+j \gamma_{2}\right).\label{eqSn1n2}
\end{equation}
The polynomial algebra can be written in the form of a deformed oscillator algebra by taking
$b^{\dagger}=I_{+}$,$b=I_{-}$, $N=K-u$, and structure function 
\begin{equation}
\Phi(N,u,H)=S_{n_{1},n_{2}}(N+u,H).
\label{structurefunction}
\end{equation} 
The constraints for finite dimensional unitary representation are known and take the form
$\Phi(0,u,E)=0$, $\Phi(p+1,u,E)=0$ and $\Phi(x,u,E)>0$ for $x=0,\ldots,p$.

\section{Dunkl oscillators}

\subsection{1D Dunkl oscillator}

The Hamiltonian of the Dunkl oscillator with one parameter is given by
\begin{equation*}
H=-\frac{1}{2}(D_{x}^{\mu})^{2}+\frac{1}{2}m^{2} x^{2},
\end{equation*}
where $m$ is some integer and the Dunkl operator given by
$D_{x}^{\mu}=\partial_{x}+\frac{\mu}{x}(1-R).$ We remark that the operator $R$, defined by the
action $R(f(x)) = f(-x),$ satisfies $R^2=I$ and
so, in the context of our presentation so far, can be viewed as the generator of the cyclic group of
order 2. The square of the Dunkl operator evaluates to
\begin{equation*}
(D_{x}^{\mu})^{2}=\partial_{x}^{2}+\frac{2\mu}{x}\partial_{x}-\frac{\mu}{x^{2}}(1-R)
\end{equation*}
We can introduce the following operators 
\begin{align}
a^{\dagger}& =\frac{1}{\sqrt{2}}(m x + D_{x}^{\mu}), 
\label{definea1}\\
a&=\frac{1}{\sqrt{2}}(m x - D_{x}^{\mu})
\label{definea2} 
\end{align}
that generate the following algebraic relations \cite{gen1,gen2}
\begin{equation*}
a^{\dagger}a=H+\frac{m}{2}(-1 - 2\mu R),\quad aa^{\dagger}=H+\frac{m}{2}(1 + 2\mu R),
\end{equation*}
\begin{equation*}
[H,a^{\dagger}]=m a^{\dagger},\quad [H,a]=-m a.
\end{equation*}
Defining
\begin{equation*}
B = \frac{a^2}{2},\  \ B^{\dagger}= \frac{(a^{\dagger})^{2}}{2},
\end{equation*}
these can be shown to satisfy the following commutation relations
\begin{equation*}
[H,B]=-2m B,\ \ [H,B^{\dagger}]= 2m B^{\dagger},
\end{equation*}
\begin{equation*}
\Rightarrow \ \ B^{\dagger}B=G(H),\quad BB^{\dagger}=G(H+2m),
\end{equation*}
where
\begin{equation*}
G(H)=\frac{1}{4}\left(H-\frac{3}{2}m+m\mu R\right)\left(H-\frac{1}{2}m-m\mu R\right).
\end{equation*}
The fact that there is a quadratic expression of the Hamiltonian is related to the existence
of two sectors unlike the case of the standard harmonic oscillator.

\subsection{2D anisotropic Dunkl oscillator}

We now introduce the anisotropic Dunkl oscillator in two dimensions:
\begin{equation*}
H=H_{x}+H_{y}=-\frac{1}{2} ({D_{x}^{\mu_x}})^{2}-\frac{1}{2} ({D_{y}^{\mu_y}})^{2}
+\frac{1}{2}m^{2}x^{2}+\frac{1}{2}n^{2}y^{2}.
\end{equation*}
For both coordinate axes, we have two polynomial Heisenberg algebras involving the generators
$R_{x}$ and $R_{y}$ (respectively) that are central elements of the algebra with
$n_{1}\gamma_{x}=n_{2}\gamma_{y}=\gamma$ (using the notation of equations (\ref{Rrels})). 
In particular, we have $\gamma_{x}=2m$, $\gamma_{y}=2n$,
$n_{1}=n$, $n_{2}=m$, $\gamma=2mn$. However, this differs from the usual anisotropic case as the
analysis of finite dimensional unitary representations will involve different sectors due to the
presence of the reflection operators $R_{x}$ and $R_{y}$ in the structure function. 

By contrast to the 1D case, we set
\begin{align}
a_x^{\dagger}& =\frac{1}{\sqrt{2}}(m x + D_{x}^{\mu_x}), 
\ \ a_y^{\dagger} =\frac{1}{\sqrt{2}}(n y + D_{y}^{\mu_y}), \label{ax}
\\
a_x&=\frac{1}{\sqrt{2}}(m x - D_{x}^{\mu_x})
\ \ a_y=\frac{1}{\sqrt{2}}(n y - D_{y}^{\mu_y})  \label{ay}
\end{align}
and also
\begin{equation*}
B_k = \frac{a_k^2}{2},\  \ B_k^{\dagger}= \frac{(a_k^{\dagger})^{2}}{2}, \ \ k=x,y
\end{equation*}
so that
\begin{align*}
& [H_x,B_x] = -2mB_x,\ \ [H_x,B_x^\dagger] =2mB_x^\dagger,\\
& [H_y,B_y] = -2nB_y,\ \ [H_y,B_y^\dagger] =2nB_y^\dagger.
\end{align*}

Making use of the generators $I_\pm$ defined in (\ref{defI}), namely
\begin{equation*}
I_- = B_x^n(B_y^{\dagger})^m,\ \ I_+ = (B_x^{\dagger})^nB_y^m,
\end{equation*}
the structure function of the 2D Dunkl
oscillator and algebraic derivation of the energy spectrum is determined by
\begin{align*}
\Phi_{HO} &=\prod_{i=1}^{n}\left[ \frac{1}{4}\left(\frac{E}{2}+2mn (x+u)- (n-i) 2m -\frac{3}{2}m + m\mu_{x}
s_{x}\right) \right.\\
& \ \ \times \left. \left(\frac{E}{2}+2mn (x+u)- (n-i) 2m -\frac{1}{2}m - m\mu_{x}
s_{x}\right)\right]\\
& \ \ \times \prod_{j=1}^{m}\left[\frac{1}{4}\left(\frac{E}{2}-2mn (x+u)+2jn  -\frac{3}{2}n + n\mu_{y}
s_{y}\right)\right.\\
& \ \ \times \left.\left(\frac{E}{2}-2mn (x+u)+2jn -\frac{1}{2}n - n\mu_{y} s_{y}\right)\right],
\end{align*}
where equations (\ref{eqSn1n2})-(\ref{structurefunction}) have been used, $s_x$ and $s_y$ are the eigenvalues of $R_x$ and $R_y$ respectively, taking values 
$1$ or $-1$.
We have two types of solutions when we impose $\Phi(0,u,E)=0$:
\begin{equation*}
u=-\frac{1}{2mn}\left(  \frac{E}{2} - (n-k_{1})2m-m -\frac{\epsilon_{1}}{2}m + \epsilon_{1} m
\mu_{x} s_{x}\right),
\end{equation*}
with the two cases corresponding to the choice $\epsilon_{1}=\{1,-1\}$ and where $k_{1}=1,\ldots,n$.
Substituting these expressions we obtain the energies using $\Phi(p+1,u,E)=0$:
\begin{equation*}
E=2mn(p+1) +m+n+\frac{\epsilon_{1}}{2}m +\frac{\epsilon_{2}}{2}n-(\epsilon_{1}m \mu_{x} s_{x}+\epsilon_{2}n \mu_{y} s_{y})+2 (mn -k_{2}n -k_{1}m),
\end{equation*}
with the corresponding structure functions
\begin{align*}
&\Phi_{HO}(x,p)=\prod_{i=1}^{n}\frac{1}{4}(2 mn x + (i-k_{1})2m)(2 mn x +(i-k_{1})2m +m
+(1-\epsilon_{1})\mu_{x}s_{x})\\
&\times \prod_{j=1}^{m}\frac{1}{4}( 2 mn (p+1) -2 mn x + 2 n(j-k_{2}))( 2 mn (p+1) -2 mn x + 2 n(j-k_{2})+n- 2\epsilon_{2} n\mu_{y}s_{y}),
\end{align*}
where $k_{1}=1,\ldots,n$ and $k_{2}=1,\ldots,m$ and $\epsilon_1,\epsilon_2\in\{1,-1\}$. It can be
verified that $\Phi_{HO}(x,p)>0$ for $x=1,\ldots,p$, which ensures that the %structure functions provide that 
the finite dimensional representations are unitary.

\section{Singular Dunkl oscillators}

\subsection{1D singular Dunkl oscillator}

The Hamiltonian of the singular Dunkl oscillator with three parameters is given by
\begin{equation*}
H=-\frac{1}{2}({D_{x}^{\mu}})^{2} +\frac{1}{2}m^{2}x^{2}+\frac{\alpha +\beta R}{2 x^{2}}.
\end{equation*}
with the Dunkl operator ${D_{x}^{\mu}}$ defined as before. In this case we introduce the following
operators which can be considered deformations of those defined in (\ref{Bdef}), and make use of
the operators $a$ and $a^\dagger$ defined in (\ref{definea1})-(\ref{definea2}) \cite{gen3}: 
\begin{equation*}
B^{\dagger}=(a^{\dagger})^{2}- \frac{\alpha+\beta R}{2 x^{2}},\quad B=a^{2}- \frac{\alpha+\beta R}{2 x^{2}}.
\end{equation*}
These operators satisfy $[B^{\dagger},R]=[B,R]=0$, and moreover
\begin{equation*}
[H,B] = -2mB,\ \ [H,B^{\dagger}]= 2m B^{\dagger}
\end{equation*}
\begin{equation}
\Rightarrow B^{\dagger}B=  H^{2}-2m H + m^{2}\left( \frac{3}{4}+\mu R - \mu^{2}- (\alpha +\beta R)\right)= G(H),
\label{gh}
\end{equation}
\begin{equation*}
BB^{\dagger}= G(H+2m).
\end{equation*}
The expression for $G(H)$ given in (\ref{gh}) can be factorised to the form
$$
G(H) = (H-\Lambda_+)(H-\Lambda_-),
$$
with 
\begin{equation}
\Lambda_\pm = m\left(1\pm\frac12\sqrt{ 1+4\alpha+4\beta s - 4\mu s+4\mu^2 s }\right),
\label{Lambdas}
\end{equation}
where $s=-1,1$ is the eigenvalue of the operator $R$.

We see that this case corresponds to a nontrivial deformation (provided $\alpha,\beta\neq
0$) of the regular 1D Dunkl oscillator introduced earlier, so the situation is genuinely different
and warrants further study. To this end, we now look at the singular Dunkl oscillator in two dimensions.

\subsection{2D singular anisotropic Dunkl oscillator}

The hamiltonian we consider consider in this section can be expressed as
\begin{align*}
H&=H_{x}+H_{y} \\
&=-\frac{1}{2} D_{x}^{\mu_x}-\frac{1}{2} D_{y}^{\mu_y} + \frac{1}{2} m^{2} x^{2} +\frac{\alpha_{x} +\beta_{x} R_{x}}{ 2 x^{2}}
  + \frac{1}{2} n^{2} y^{2} +\frac{\alpha_{y} +\beta_{y} R_{y}}{ 2 y^{2}}.
\end{align*}

We have the following two-polynomial Heisenberg algebras that contain the grading elements $R_{x}$ and
$R_{y}$. We have similar integrals and deformed oscillator form with $n_{1}=n, n_{2}=m$,
$\gamma_{x}=2m$ , $\gamma_{y}=2n$, $\gamma= mn$. Specifically, making use of the operators 
$a_x,a_x^\dagger,a_y,a_y^\dagger$ given in (\ref{ax})-(\ref{ay}), we set
\begin{align*}
& B_x = {a_x^2}- \frac{\alpha_x+\beta_x R_x}{2 x^{2}},\  \ 
B_x^{\dagger}= (a_x^{\dagger})^{2}- \frac{\alpha_x+\beta_x R_x}{2 x^{2}},\\
& B_y = {a_y^2}- \frac{\alpha_y+\beta_y R_y}{2 y^{2}},\  \ 
B_y^{\dagger}= {(a_y^{\dagger})^{2}}- \frac{\alpha_y+\beta_y R_y}{2 y^{2}},
\end{align*}
with
\begin{align*}
& [H_x,B_x] = -2mB_x,\ \ [H_x,B_x^\dagger] =2mB_x^\dagger,\\
& [H_y,B_y] = -2nB_y,\ \ [H_y,B_y^\dagger] =2nB_y^\dagger.
\end{align*}
The structure function (again making use of
the generators of the form (\ref{defI})) for the singular oscillator can be written as
\begin{align*}
\Phi_{SHO}&=\prod_{i=1}^{n}\left[\left(\frac{E}{2}+2mn(x+u)-(n-i)2m
-m\left(1-\frac{s_{x}}{2}+\nu_{x,s_{x}}\right)\right) \right. \\
&\ \ \times \left.\left(\frac{E}{2}+2mn(x+u)-(n-i)2m -m\left(1+\frac{s_{x}}{2}-\nu_{x,s_{x}}\right)\right)\right]  \\
&\ \ \times \prod_{j=1}^{m}\left[ \left(\frac{E}{2}-2mn(x+u)+2jn
-n\left(1-\frac{s_{y}}{2}+\nu_{y,s_{y}}\right)\right)
\right.\\
& \ \ \times \left. \left(\frac{E}{2}-2mn(x+u)+2jn -n\left(1+\frac{s_{y}}{2}-\nu_{y,s_{y}}\right)\right)\right].
\end{align*}
As before, equations (\ref{eqSn1n2})-(\ref{structurefunction}) have been used, $s_x$ and $s_y$
represent eigenvalues of the operators $R_x$ and $R_y$ respectively, and so take on values $1$ or
$-1$. To avoid overly complicated expressions, particularly arising from the form of $G(H)$ as
given in (\ref{gh}), we have also introduced the notation 
$$
\nu_{x,s_x}  =   \frac{s_x}{2} - \frac12\sqrt{1+4\alpha_x+4\beta_x s_x - 4\mu_x s_x+4\mu_x^2 s_x}
$$
(and a similar expression for $\nu_{y,s_y}$), the form of which has been chosen for its connection
to the physical spectrum. This last point shall be discussed in Section \ref{finalsec} below.

Using $\Phi(0,u,E)=0$, we obtain
\begin{equation*}
u=\frac{1}{2mn}\left(-\frac{E}{2}+(n-k_{1})2m +
m\left(1-\epsilon_{1}\frac{s_{x}}{2}+\epsilon_{1}\nu_{x,s_{x}}\right)\right).
\end{equation*}
Applying the condition $\Phi(p+1,u,E)=0$ determines the energy spectrum as
\begin{equation*}
E=2mn (p+1) +2 (nm -k_{2}n -k_{1}m)+m
\left(1-\epsilon_{1}\frac{s_{x}}{2}+\epsilon_{1}\nu_{x,s_{x}}\right)+n\left(1-\epsilon_{2}\frac{s_{y}}{2}+\epsilon_{2}\nu_{y,s_{y}}\right).
\end{equation*}
Once again, we have made use of the notation $\epsilon_1,\epsilon_2=1,-1$ to characterize the
various cases that arise in this Daskaloyannis approach.
The associated structure functions reduce to the form
\begin{align*}
& \Phi_{SHO}(x,p)\\
&\ \ =\prod_{i=1}^{n}(2 mn x +( i-k_{1})2m) (2mn x +(i-k_{1})2m +m
(-\epsilon_{1}s_{x}+2\epsilon_{1}\nu_{x,s_{x}})) \\
&\ \ \times  \prod_{j=1}^{m}( 2 mn (p+1) - 2 mn x  + 2 n(j-k_{2}))
(2mn (p+1) -2mnx +2n (j-k_{2})+n (-\epsilon_{2}s_{y}+2\epsilon_{2}\nu_{y,s_{y}})).
\end{align*}
where $k_{1}=1,\ldots,n$ and $k_{2}=1,\ldots,m.$. It can be verified also for this structure
function that $\Phi_{HO}(x,p)>0$ for $x=1,\ldots,p.$

\section{A comment on the physical spectrum} \label{finalsec}

The physical spectrum of these two new 2D anisotropic Dunkl and singular Dunkl oscillators can be
obtained using separation of variables and explicit expressions for the wavefunctions in the 1D
cases \cite{gen1,gen2,gen3}. In the Dunkl case there are four different sectors, characterised by
the eigenvalues of $R_x$ and $R_y$: 
\begin{equation*}
E=(2 m n_{1x}+2n n_{1y}+m(\mu_{x}+1-\frac{1}{2}s_x)+n(\mu_{y}+1-\frac{1}{2}s_y)).
\end{equation*}
In the case of the 2D singular oscillator spectrum we also have four sectors
\begin{equation*}
E=( 2 m n_{1x} + 2 n n_{1y} + m(\nu_{x,+}+1-\frac{1}{2}s_x)+n(\nu_{y,+}+1-\frac{1}{2}s_y)),
\end{equation*}
with
\begin{equation*}
\nu_{i,\pm}= 2 k_{i}^{\pm} +\mu_{i},\quad i=x,y
\end{equation*}
and
\[ \nu_{i,+}+\frac{1}{2}>0,\quad  \nu_{i,-}+\frac{3}{2}>0. \]
The parameters $k_i^\pm$ of the 2D singular oscillator must satisfy
\begin{equation*}
\alpha_{i}= 2 k_{i}^{+}(k_{i}^{+} +\mu_{i} -\frac{1}{2}) + 2 k_{i}^{-} ( k_{i}^{-}+\mu_{i}
+\frac{1}{2}),
\end{equation*}
\begin{equation*}
\beta_{i}= 2 k_{i}^{+}(k_{i}^{+} +\mu_{i} -\frac{1}{2}) - 2 k_{i}^{-} ( k_{i}^{-}+\mu_{i}
+\frac{1}{2}).
\end{equation*}
It turns out that not all eigenvalues arising from the algebraic approach are physical. In order to
compare the physical spectrum and those eigenvalues obtained by the algebraic derivation via
deformed oscillator algebra, we define
\begin{equation}
n_{1x}'=n n_{1x} + l_{1}, n_{2x}'=n n_{2x} + l_{1}
\end{equation}
\begin{equation}
n_{1y}'=m n_{1y} + l_{2}, n_{2y}'=m n_{2y} + l_{2}
\end{equation}
with $l_{1}=0,\ldots,n-1$ and  $l_{2}=0,\ldots,m-1$ and 
$n_{1x},n_{2x},n_{1y},n_{2y}=0,1,2,\ldots$ and we take respectively for the four physical solutions
$p= n_{ix}'+n_{jy}'$ for $i,j=1,2$ and also considering $l_{1}=-k_{1}+n$ and $l_{2}=-k_{2}+m$

 %
 %===========================================================================
 %
 \section{Conclusion}

In this paper, we have extended the construction of integrals from substructure as ladder, shift or supercharges used in recent years to introduce several new families of superintegrable systems to ladder operators involving generator of the cyclic group and also generalized the Daskaloyannis approach for the corresponding class of polynomial algebras. We applied this construction for a sum of two $c_{\lambda}$ extended oscillators.  \par
%
%---
% 
We also introduced two new families of superintegrable systems that are the anisotropic Dunkl and
singular Dunkl oscillators for which we demonstrated how to apply the general construction. We compare the spectrum that divide into four sector to the physical spectrum derived by separation of variables. \par
%
%---
%
Superintegrable systems with higher order integrals of motion with Dunkl operator or more generally
grading are unexplored and no systematic study has been undertaken. \par

 \setcounter{equation}{0}

\par
%
%============================================================================
%
\section*{ACKNOWLEDGEMENTS}

The research of I.\ M.\ was supported by the Australian Research Council through Discovery Project No.\ DP110101414 and Discovery Early Career Researcher Award DE130101067.\par
%
%============================================================================
%

%
%======================================================================
%
%\newpage

\end{document}